\documentclass[aps,showpacs,tightenlines,nobalancelastpage,floatfix,
  twocolumn]{revtex4}

\usepackage{graphicx}
\usepackage{amsmath}
\usepackage{amsfonts}
\usepackage{amssymb}
\usepackage{bbm}
\usepackage{epic}
\usepackage{bm}

\begin{document}

\title{Time-optimal synthesis of $SU(2)$ transformations for a
  spin-$1/2$ system}

\date{\today}

\author{A. D. Boozer}

\email{boozer@unm.edu}

\affiliation{
  Department of Physics,
  University of New Mexico
  Albuquerque, New Mexico 87131
}

\begin{abstract}
  We consider a quantum control problem involving a spin-$1/2$
  particle in a magnetic field.
  The magnitude of the field is held constant, and the direction of
  the field,
  which is constrained to lie in the $x-y$ plane, serves as a control
  parameter that can be varied to govern the evolution of the system.
  We analytically solve for the time dependence of the control
  parameter that will synthesize a given target $SU(2)$ transformation
  in the least possible amount of time, and we show that the
  time-optimal solutions have a simple geometric interpretation in
  terms of the fiber bundle structure of $SU(2)$.
  We also generalize our time-optimal solutions to a control problem
  that includes a constant bias field along the $\bm{\hat{z}}$ axis,
  and to the case of inhomogeneous control, in which a single control
  parameter governs the evolution of an ensemble of spin-$1/2$
  systems.
\end{abstract}

\pacs{
  42.50.Dv,   % Quantum state engineering and measurements 
  02.30.Yy,   % Control theory 
  02.20.Sv    % Lie algebras of Lie groups
}

\maketitle

\section{Introduction}

Many applications rely on the ability to coherently control the state
of a quantum system
\cite{shapiro, judson, brumer, rabitz-2000, khaneja, ramanathan}.
In particular, the current push to develop robust quantum information
processors has led to the development of quantum control protocols for
a diverse array of experimental platforms, including atomic, optical,
and condensed matter systems \cite{viola, grace, chiara, nielsen}.
In a typical control problem, the system in question is described by
a Hamiltonian containing several control parameters that we are free
to vary, and we would like to determine the time dependence of these
parameters such that the evolution of the system implements a desired
unitary transformation.
Such problems are generally highly nontrivial: they do not usually
admit an analytic solution, and must be solved via numerical searches
\cite{dahleh, rabitz, shen, hsieh, moore, walmsley, tesch, palao}.
Analytic solutions can, however, sometimes be obtained for control
problems involving low-dimensional systems.
In particular, for control problems involving a spin-$1/2$ particle,
analytic solutions have been obtained that minimize either an
energy-type cost
functional \cite{boscain, dalessandro-2000, dalessandro} or the total
evolution time \cite{khaneja-2001, boscain-2006}.

Here we consider a model quantum control problem involving a
spin-$1/2$ particle in a magnetic field.
The magnitude of the field is held constant, and its direction, which
is constrained to lie in the $x-y$ plane, serves as a control
parameter that can be varied to govern the evolution of the system.
The evolution can be described in terms of an $SU(2)$ evolution
operator $U(t)$, such that if the state of the spin at time zero is
$|\psi(0)\rangle$ then the state at time $t$ is
$|\psi(t)\rangle = U(t)|\psi(0)\rangle$.
Given an arbitrary target $SU(2)$ transformation $V$, we analytically
solve for the time dependence of the control parameter such that
$U(t) = V$ and $t$ is as small as possible.
By viewing $SU(2)$ as a $U(1)$ fiber bundle over the two-dimensional
sphere $S^2$, we are able to give a simple geometric interpretation to
these time-optimal solutions.
We also generalize our time-optimal solutions to a control problem
that includes a constant bias field along the $\bm{\hat{z}}$ axis.

An important development in the field of quantum control is the notion
of inhomogeneous control, in which a single set of control parameters
governs the evolution of an ensemble of systems subject to different
Hamiltonians.
The differences in the Hamiltonians may, for example, describe
unwanted perturbations that give rise to decoherence.
By choosing the control parameters properly, one can compensate for
these perturbations so that the resulting system dynamics are
insensitive to their presence \cite{vandersypen, cummins}.
Alternatively, the differences in the Hamiltonians may be intentional,
so as to provide a means of addressing individual systems in the
ensemble \cite{li, kobzar, khaneja-2005}.

We investigate inhomogeneous control in our model control problem by
generalizing the problem to the case of an ensemble of $N$ spin-$1/2$
systems.
The magnetic fields of the different systems vary in magnitude but are
all aligned along a common direction in the $x-y$ plane, and we take
this common direction to be the control parameter that governs the
evolution of the entire ensemble.
We obtain a semi-analytic solution to this inhomogeneous control
problem for the case $N=2$, and we verify that our solution is
time-optimal by comparing it with the results of a numerical search.

\section{Control problem}
\label{sec:model-control-problem}

The system that we consider consists of a spin-$1/2$ particle in a
magnetic field $\bm{B}$.
We assume that the magnitude $B \equiv |\bm{B}|$ of the magnetic
field is constant, and its direction
$\bm{\hat{n}} \equiv \bm{B}/|\bm{B}|$ serves as a control parameter
that can be varied to govern the evolution of the system.
The Hamiltonian for the system is
\begin{align}
  \label{eqn:hamiltonian}
  H = -\mu B \bm{\sigma}\cdot\bm{\hat{n}},
\end{align}
where $\mu$ is the magnetic moment of the particle and $\sigma_k$ are
the Pauli spin matrices.
For simplicity, we will choose units such that $\mu B = 1$.
The system evolves in time according to the unitary transformation
\begin{eqnarray}
  \label{eqn:U}
  U(t) = T\exp(-i\int_0^t H(t')\,dt'),
\end{eqnarray}
where $T$ is a time-ordering operator that places operators at early
times to the right of operators at later times.
We note that $U$ satisfies the Schr\"{o}dinger equation
\begin{align}
  \label{eqn:schrodinger-eqn}
  i\dot{U} = HU.
\end{align}
From Eq.~(\ref{eqn:U}), and the fact that $H$ is traceless, it follows
that $\det U = 1$, so $U$ is an $SU(2)$ transformation.

We now consider a control problem in which we are given a target
$SU(2)$ transformation $V$ and are asked to determine the
time dependence of the control parameter $\bm{\hat{n}}$ and the total
evolution time $t$ such that $U(t) = V$ and $t$ is as small as
possible.
If $\bm{\hat{n}}$ is allowed to point in any direction, then the
solution to the control problem is trivial: we write $V$ in the form
$V = e^{i\bm{r}\cdot\bm{\sigma}}$, where $|\bm{r}| \leq \pi$, and
we take
\begin{align}
  \label{eqn:unconstrained-solution}
  \bm{\hat{n}} &= \bm{\hat{r}}, &
  t &= |\bm{r}|.
\end{align}
For example, for a target transformation $V = e^{i\eta\sigma_z/2}$
describing a spatial rotation with axis $\bm{\hat{z}}$ and angle
$\eta$, we find that $\bm{\hat{n}} = \bm{\hat{z}}$ and $t = \eta/2$.

Let us suppose, however, that the control parameter $\bm{\hat{n}}$ is
constrained to lie in the $x-y$ plane.
The control problem is still solvable, but the solution is no longer
trivial.
We can verify that the control problem is solvable by presenting a
solution that is not time-optimal.
Let us write the target transformation $V$ in terms of Euler
angles $\psi$, $\theta$, and $\phi$:
\begin{align}
  \label{eqn:euler-solution}
  V = e^{i\psi\sigma_x/2} e^{i\theta\sigma_y/2} e^{i\phi\sigma_x/2}.
\end{align}
From Eq.~(\ref{eqn:euler-solution}), it follows that $V$ can be
synthesized by by taking
\begin{align}
  \label{eqn:euler-solution-n}
  \bm{\hat{n}}(\tau) &=
  \left\{
  \begin{array}{cl}
    \bm{\hat{x}} &
    \mbox{for $0 < \tau < |\phi/2|$}, \\
    \bm{\hat{y}} &
    \mbox{for $|\phi/2| < \tau < |\phi/2| + |\theta/2|$}, \\
    \bm{\hat{x}} &
    \mbox{for $|\phi/2| + |\theta/2| < \tau < t$}, \\
  \end{array}
  \right. \\
  \label{eqn:euler-solution-t}
  t &= |\psi/2| + |\theta/2| + |\phi/2|.
\end{align}
For example, consider again a target transformation
$V = e^{i\eta\sigma_z/2}$ describing a spatial rotation with axis
$\bm{\hat{z}}$ and angle $\eta$.
We find that
$V = e^{-i\pi\sigma_x/4} e^{i\eta\sigma_y/2} e^{i\pi\sigma_x/4}$, so
$\phi = -\psi = \pi/2$, $\theta = \eta$, and $t = \pi/2 + \eta/2$.
For comparison, recall that $t = \eta/2$ for the unconstrained
control problem in which $\bm{\hat{n}}$ is allowed to point in any
direction.

\section{Time-optimal solution}
\label{sec:homogeneous-problem}

We now present a time-optimal solution to the constrained control
problem.
We begin by describing two methods for assigning coordinates to an
$SU(2)$ transformation $U$.
For the first method, we assign real-valued coordinates
$\bm{r} = (w, x, y, z)$ to $U$ by expanding $U$ in the Pauli spin
matrices:
\begin{align}
  \label{eqn:coordinates-embedding}
  U = w + i x\sigma_x + i y\sigma_y + i z\sigma_z.
\end{align}
We call these coordinates embedding coordinates, because they describe
an embedding of $SU(2)$ into $\mathbb{R}^4$.
For the second method, we assign complex-valued coordinates
$(z_1, z_2)$ to $U$ by expressing $U$ in the form
\begin{align}
  \label{eqn:coordinates-complex}
  U =
  \left(
  \begin{array}{cc}
    z_1 & z_2 \\
    -z_2^* & z_1^* \\
  \end{array}
  \right).
\end{align}
We call these coordinates complex coordinates.
From Eqs.~(\ref{eqn:coordinates-embedding}) and
(\ref{eqn:coordinates-complex}), it follows that the two sets of
coordinates are related by $(z_1, z_2) = (w + i z, y + ix)$.

The Lie group $SU(2)$ is three-dimensional, but both sets of
coordinates label $SU(2)$ transformations using four real parameters.
So for both sets of coordinates there are more coordinate degrees of
freedom than physical degrees of freedom, and only some of the points
in the coordinate space actually correspond to $SU(2)$
transformations.
From Eqs.~(\ref{eqn:coordinates-embedding}) and
(\ref{eqn:coordinates-complex}), it follows that such points satisfy
the constraint
\begin{align}
  \label{eqn:constraint}
  |\bm{r}|^2 = |z_1|^2 + |z_2|^2 = 1.
\end{align}
The locus of points $\bm{r}$ that satisfy Eq.~(\ref{eqn:constraint})
is a three-dimensional sphere $S^3$ embedded in $\mathbbm{R}^4$, and
the mapping $U \mapsto \bm{r}$ is a diffeomorphism from $SU(2)$ to
$S^3$.

It is useful to express the Schr\"{o}dinger equation
(\ref{eqn:schrodinger-eqn}) in terms of both sets of coordinates.
We first consider the embedding coordinates.
We substitute the definition of the embedding coordinates given in
Eq.~(\ref{eqn:coordinates-embedding}) into the Schr\"{o}dinger
equation (\ref{eqn:schrodinger-eqn}) to obtain an equation of motion
for $\bm{r}$:
\begin{align}
  \label{eqn:schrodinger-embedding}
  \dot{\bm{r}} =
  n_x \bm{L}_x(\bm{r}) +
  n_y \bm{L}_y(\bm{r}) +
  n_z \bm{L}_z(\bm{r}),
\end{align}
where
\begin{align}
  \bm{L}_x(\bm{r}) &=
  w\bm{\hat{x}} - x\bm{\hat{w}} + z\bm{\hat{y}} - y\bm{\hat{z}}, \\
  \bm{L}_y(\bm{r}) &=
  w\bm{\hat{y}} - y\bm{\hat{w}} + x\bm{\hat{z}} - z\bm{\hat{x}}, \\
  \label{eqn:Lz}
  \bm{L}_z(\bm{r}) &=
  w\bm{\hat{z}} - z\bm{\hat{w}} + y\bm{\hat{x}} - x\bm{\hat{y}},
\end{align}
are orthonormal vectors that span the tangent space of $S^3$ at the
point $\bm{r}$.
As $\bm{r}$ evolves in time, it traces out a path in $S^3$ whose
tangent vector is $\dot{\bm{r}}$.
From Eq.~(\ref{eqn:schrodinger-embedding}) it follows that the length
of the tangent vector is $|\dot{\bm{r}}| = 1$, so time corresponds to
arc length along the path.
The time evolution of $\bm{r}$ is governed by the control parameter
$\bm{\hat{n}}$, which dictates the projection of the tangent vector
$\dot{\bm{r}}$ along the basis vectors $\bm{L}_k(\bm{r})$:
\begin{align}
  \bm{L}_x(\bm{r}) \cdot \dot{\bm{r}} &= n_x, \\
  \bm{L}_y(\bm{r}) \cdot \dot{\bm{r}} &= n_y, \\
  \bm{L}_z(\bm{r}) \cdot \dot{\bm{r}} &= n_z.
\end{align}
For the constrained control problem
$\bm{\hat{n}} = \cos\phi\,\bm{\hat{x}} + \sin\phi\,\bm{\hat{y}}$ for
some angle $\phi$, so
\begin{align}
  \label{eqn:projection-x}
  \bm{L}_x(\bm{r}) \cdot \dot{\bm{r}} &= \cos\phi, \\
  \label{eqn:projection-y}
  \bm{L}_y(\bm{r}) \cdot \dot{\bm{r}} &= \sin\phi, \\
  \label{eqn:projection-z}
  \bm{L}_z(\bm{r}) \cdot \dot{\bm{r}} &= 0.
\end{align}

It is also useful to express the Schr\"{o}dinger equation
(\ref{eqn:schrodinger-eqn}) in terms of the complex coordinates.
We substitute the definition of the complex coordinates given in
Eq.~(\ref{eqn:coordinates-complex}) into
Eq.~(\ref{eqn:schrodinger-eqn}) to obtain equations of motion for
$z_1$ and $z_2$:
\begin{align}
  \label{eqn:schrodinger-complex-a}
  \dot{z}_1 &= -i e^{-i\phi}\,z_2^*, \\
  \label{eqn:schrodinger-complex-b}
  \dot{z}_2 &= i e^{-i\phi}\,z_1^*.
\end{align}
If we differentiate Eqs.~(\ref{eqn:schrodinger-complex-a}) and
(\ref{eqn:schrodinger-complex-b}) with respect to $t$ and then
substitute for $\dot{z}_1$ and $\dot{z}_2$ using the original
equations, we obtain the decoupled equations
\begin{align} 
  \label{eqn:schrodinger-complex-decoupled-a}
  \ddot{z}_1 + i\dot{\phi}\dot{z}_1 + z_1 &= 0, \\
  \label{eqn:schrodinger-complex-decoupled-b}
  \ddot{z}_2 + i\dot{\phi}\dot{z}_2 + z_2 &= 0.
\end{align}
Using Eqs.~(\ref{eqn:schrodinger-complex-a}) and
(\ref{eqn:schrodinger-complex-b}), it is straightforward to derive the
identities
\begin{align}
  \label{eqn:complex-constraint-a}
  \dot{z}_1 z_1^* + \dot{z}_2 z_2^* &= 0, \\
  \label{eqn:complex-constraint-b}
  \dot{z}_2 z_1 - \dot{z}_1 z_2 &= i e^{-i\phi}.
\end{align}
We can understand the meaning of these identities by transforming
from complex coordinates to embedding coordinates:
\begin{align}
  \dot{z}_1 z_1^* + \dot{z}_2 z_2^* &=
  \bm{r}\cdot\dot{\bm{r}} +
  i \bm{L}_z(\bm{r})\cdot \dot{\bm{r}}, \\
  \dot{z}_2 z_1 - \dot{z}_1 z_2 &=
  \bm{L}_y(\bm{r})\cdot \dot{\bm{r}} +
  i \bm{L}_x(\bm{r})\cdot \dot{\bm{r}}.
\end{align}
So Eqs.~(\ref{eqn:complex-constraint-a}) and
(\ref{eqn:complex-constraint-b}) follow from
Eqs.~(\ref{eqn:constraint}) and
(\ref{eqn:projection-x})--(\ref{eqn:projection-z}).

Let us now return to the embedding coordinates and consider the
problem of finding a minimum-length path in $S^3$ that satisfies the
constraint $\bm{L}_z(\bm{r})\cdot \dot{\bm{r}} = 0$.
Such a path can be obtained by minimizing the action
\begin{align}
  \label{eqn:action}
  S =
  \int (|\bm{r}'| +
  \gamma(|\bm{r}|^2 - 1) +
  \lambda \bm{L}_z(\bm{r})\cdot \bm{r}')\,du.
\end{align}
Here $u$ is an arbitrary parameterization of the path,
$\bm{r}' \equiv d\bm{r}/du$, and $\gamma$ and $\lambda$ are Lagrange
multipliers.
The first term of the integrand gives the length of the path, the
second term imposes the constraint $|\bm{r}|^2 = 1$, which restricts
the path to $S^3$,
and the third term imposes the constraint
$\bm{L}_z(\bm{r}) \cdot \dot{\bm{r}} = 0$, which expresses the fact
that the control parameter $\bm{\hat{n}}$ must lie in the $x-y$ plane.
Note that $\bm{r}' \equiv d\bm{r}/du = (dt/du)\dot{\bm{r}}$ and
$|\dot{\bm{r}}| = 1$, so the parameter $u$ is related to the time $t$
by
\begin{eqnarray}
  \label{eqn:t-u}
  dt/du = |\bm{r}'|.
\end{eqnarray}
We write down the Euler-Lagrange equations corresponding to the
action given in Eq.~(\ref{eqn:action}), use Eq.~(\ref{eqn:t-u}) to
replace $u$ with $t$, and transform from embedding coordinates to
complex coordinates to obtain
\begin{align}
  \label{eqn:euler-lagrange-a}
  \ddot{z}_1 + 2i\lambda \dot{z}_1 +
  (i\dot{\lambda} - 2\gamma)z_1 &= 0, \\
  \label{eqn:euler-lagrange-b}
  \ddot{z}_2 + 2i\lambda \dot{z}_2 +
  (i\dot{\lambda} - 2\gamma) z_2 &= 0.
\end{align}

A time-optimal solution to the constrained control problem must satisfy
the Schr\"{o}dinger equations
(\ref{eqn:schrodinger-complex-decoupled-a}) and
(\ref{eqn:schrodinger-complex-decoupled-b})
as well as the Euler-Lagrange equations
(\ref{eqn:euler-lagrange-a}) and
(\ref{eqn:euler-lagrange-b}).
We subtract
Eq.~(\ref{eqn:schrodinger-complex-decoupled-a}) from
(\ref{eqn:euler-lagrange-a}) and
Eq.~(\ref{eqn:schrodinger-complex-decoupled-b}) from
(\ref{eqn:euler-lagrange-b})
to obtain
\begin{align}
  \label{eqn:subtracted-a}
  i(2\lambda - \dot{\phi})\dot{z}_1 +
  (i\dot{\lambda} - 2\gamma - 1)z_1 &= 0, \\
  \label{eqn:subtracted-b}
  i(2\lambda - \dot{\phi})\dot{z}_2 +
  (i\dot{\lambda} - 2\gamma - 1) z_2 &= 0.
\end{align}
Using the identities given in Eqs.~(\ref{eqn:complex-constraint-a})
and (\ref{eqn:complex-constraint-b}), we can eliminate the coordinates
$z_1$ and $z_2$ from Eq.~(\ref{eqn:subtracted-a}) and
(\ref{eqn:subtracted-b}) and obtain equations that involve only the
parameters $\gamma$, $\lambda$, and $\phi$:
\begin{align}
  \label{eqn:homogeneous-problem-1}
  \dot{\phi} &= 2\lambda, &
  i\dot{\lambda} &= 2\gamma + 1.
\end{align}
The solution to these equations is
\begin{align}
  \label{eqn:solution-phi}
  \gamma &= -1/2, &
  \lambda &= \omega/2, &
  \phi &= \phi_0 + \omega t,
\end{align}
where $\phi_0$ and $\omega$ are integration constants.
So an $SU(2)$ transformation can be synthesized in a time-optimal
fashion by varying the control parameter $\phi$ as described by
Eq.~(\ref{eqn:solution-phi}).

We would now like to calculate the evolution operator $U$ that results
when the control parameter $\phi$ is varied in the time-optimal
fashion described by Eq.~(\ref{eqn:solution-phi}).
We first note that $U(0)$ is the identity transformation, which has
complex coordinates $(z_1, z_2) = (1,0)$.
We substitute Eq.~(\ref{eqn:solution-phi}) for $\phi$ into the
Schr\"{o}dinger equations
(\ref{eqn:schrodinger-complex-a}) and
(\ref{eqn:schrodinger-complex-b})
and solve them subject to these initial conditions to obtain
\begin{align}
  \label{eqn:z1}
  z_1 &=
  (2\alpha)^{-1}(\beta_+ e^{i\beta_- t} + \beta_- e^{-i\beta_+ t}), \\
  \label{eqn:z2}
  z_2 &=
  (2\alpha)^{-1} e^{-i\phi_0} (e^{i\beta_- t} - e^{-i\beta_+ t}),
\end{align}
where
\begin{align}
  \label{eqn:alpha-beta}
  \alpha &\equiv (1 + \omega^2/4)^{1/2}, &
  \beta_\pm &\equiv \alpha \pm \omega/2.
\end{align}
It is useful to view the parameters $(\phi_0, \omega, t)$ as defining
a third set of coordinates for $U$.
We call these coordinates time-optimal coordinates.
Eqs.~(\ref{eqn:z1}) and (\ref{eqn:z2}) can then be viewed as
describing a coordinate transformation from time-optimal coordinates
to complex coordinates.

Suppose we are given a target $SU(2)$ transformation $V$.
We can synthesize $V$ in a time-optimal fashion by determining its
complex coordinates $(z_1, z_2)$ and then inverting
Eqs.~(\ref{eqn:z1}) and (\ref{eqn:z2}) to obtain its time-optimal
coordinates $(\phi_0, \omega, t)$.
The parameters $\phi_0$ and $\omega$ tell us the time dependence of
the control parameter $\phi$, and the parameter $t$ tells us the total
evolution time.

Let us now consider some specific examples.
First we consider a target transformation
$V = e^{i\eta \bm{\hat{e}}_\theta \cdot \bm{\sigma}/2}$ that describes
a spatial rotation with axis
$\bm{\hat{e}}_\theta \equiv
\cos\theta\,\bm{\hat{x}} + \sin\theta\,\bm{\hat{y}}$ and angle $\eta$.
The complex coordinates of $V$ are
$(z_1, z_2) = (\cos\eta/2, i e^{-i\theta}\sin\eta/2)$.
We invert Eqs.~(\ref{eqn:z1}) and (\ref{eqn:z2}) to obtain the
time-optimal coordinates:
\begin{align}
  \phi_0 &= \theta, &
  \omega &= 0, &
  t &= \eta/2.
\end{align}
This solution is identical to the time-optimal solution for the
unconstrained control problem described in
Eq.~(\ref{eqn:unconstrained-solution}).
This is to be expected, since the time-optimal solution for the
unconstrained control problem satisfies the constraint that
$\bm{\hat{n}}$ must lie in the $x-y$ plane.

\begin{figure}[t]
  \centering
  \includegraphics{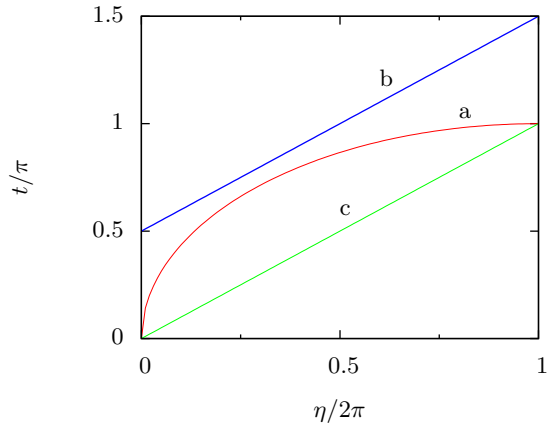}
  \caption{
    \label{fig:compare}
    (Color online)
    Time $t$ needed to synthesize the transformation
    $V = e^{i\eta\sigma_z/2}$ versus $\eta$.
    (a) Time-optimal solution for the constrained control problem.
    (b) Euler solution for the constrained control problem.
    (c) Time-optimal solution for the unconstrained control problem.
  }
\end{figure}

Next we consider a target transformation $V = e^{i\eta \sigma_z/2}$
that describes a spatial rotation with axis $\bm{\hat{z}}$ and angle
$\eta$.
The complex coordinates of $V$ are $(z_1, z_2) = (e^{i\eta/2}, 0)$.
We invert Eqs.~(\ref{eqn:z1}) and (\ref{eqn:z2}) to obtain the
time-optimal coordinates:
\begin{align}
  \label{eqn:solution-z-rotation}
  \omega &= 2\nu(1 - \nu^2)^{-1/2}, &
  t &= \pi(1 - \nu^2)^{1/2},
\end{align}
where $\nu \equiv 1 - \eta/2\pi$.
The parameter $\phi_0$ is undetermined by the inversion, and any value
can be used to perform a time-optimal synthesis of $V$.
Mathematically, $\phi_0$ is undetermined because $V$ is located at a
coordinate singularity of the time-optimal coordinate system;
physically, it is because $V$ is invariant under similarity
transformations involving arbitrary rotations about the $\bm{\hat{z}}$
axis.
In Fig.~\ref{fig:compare} we compare the time-optimal solution
described in Eq.~(\ref{eqn:solution-z-rotation}) with the Euler
solution described in
Eqs.~(\ref{eqn:euler-solution-n})--(\ref{eqn:euler-solution-t}) and
the time-optimal solution for the unconstrained control problem
described in Eq.~(\ref{eqn:unconstrained-solution}).

\begin{figure}[t]
  \centering
  \includegraphics{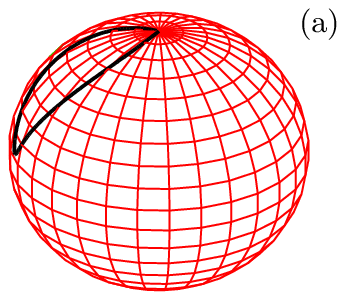}
  \includegraphics{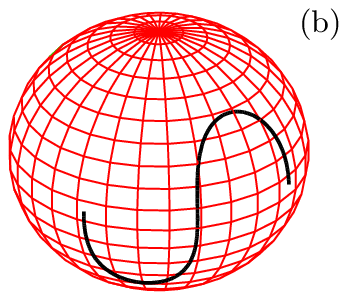}
  \caption{
    \label{fig:spin-trajectory}
    (Color online)
    Trajectory of the spin on the Bloch sphere for the time-optimal
    synthesis of the transformation $V = e^{i\pi\sigma_z/4}$, which
    describes $\pi/2$-rotation about the $\bm{\hat{z}}$ axis.
    (a) Spin initially aligned along the $\bm{\hat{z}}$ axis.
    (b) Spin initially aligned along the $-\bm{\hat{y}}$ axis.
  }
\end{figure}

Let us now consider the trajectory of the spin on the Bloch sphere as
it evolves along a time-optimal path.
If the state of the spin at time zero is $|\psi(0)\rangle$, then the
state at time $t$ is $|\psi(t)\rangle = U(t)|\psi(0)\rangle$.
We can represent the state of the spin at time $t$ as a point
$\bm{\hat{s}}(t) = \langle \psi(t) | \bm{\sigma} |\psi(t)\rangle$
on the Bloch sphere.
In Fig.~\ref{fig:spin-trajectory} we plot the trajectory of the spin
on the Bloch sphere for the time-optimal synthesis of a
$\pi/2$-rotation about the $\bm{\hat{z}}$ axis
($V = e^{i\pi\sigma_z/4}$),
where the spin is initially aligned along the $\bm{\hat{z}}$ axis for
Fig.~\ref{fig:spin-trajectory}(a) and the $-\bm{\hat{y}}$ axis for
Fig.~\ref{fig:spin-trajectory}(b).
For both curves we take $\phi_0 = 0$.

\section{Properties of the time-optimal solutions}
\label{sec:properties}

We can visualize the time-optimal solutions by representing $SU(2)$
transformations as points on the two-dimensional sphere $S^2$.
Given an $SU(2)$ transformation $U$, we define $\bm{\hat{p}}(U)$ to be
the point on $S^2$ corresponding to the state
$U^\dagger|\uparrow\rangle$:
\begin{align}
  \label{eqn:hopf}
  \bm{\hat{p}}(U) =
  \langle \uparrow | U \bm{\sigma} U^\dagger | \uparrow \rangle.
\end{align}
We note that
$\bm{\hat{p}}(U) = \bm{\hat{p}}(e^{i\theta\sigma_z} U)$ for any value
of $\theta$.
This property of $\bm{\hat{p}}$ allows us to view $SU(2)$ as a fiber
bundle, where $S^2$ is the base manifold, $U(1)$ is the fiber, and
$\bm{\hat{p}}:SU(2) \rightarrow S^2$ is the projection function.

\begin{figure}[t]
  \centering
  \includegraphics{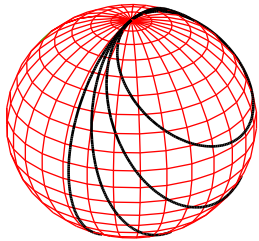}
  \caption{
    \label{projection}
    (Color online)
    Paths on the two-dimensional sphere $S^2$ for the time-optimal
    synthesis of the transformation $V = e^{i\eta\sigma_z/2}$, where
    $\eta = \pi/2, \pi, 3\pi/2, 2\pi$.
    Longer paths correspond to larger values of $\eta$.
  }
\end{figure}

We will now show that the time-optimal solutions project to circles on
$S^2$.
Let us identify the plane that bisects $S^2$ at the equator with the
complex plane.
We can map points $\bm{\hat{p}}$ on $S^2$ to complex numbers
$\zeta(\bm{\hat{p}})$ on the complex plane by stereographically
projecting from the south pole:
\begin{align}
  \label{eqn:stereographic}
  \zeta(\bm{\hat{p}}) = \frac{p_x + i p_y}{1 + p_z}.
\end{align}
Let $(z_1, z_2)$ denote the complex coordinates of an arbitrary
$SU(2)$ transformation $U$.
From Eqs.~(\ref{eqn:coordinates-complex}), (\ref{eqn:hopf}), and
(\ref{eqn:stereographic}), it follows that
\begin{align}
  \label{eqn:stereographic-hopf}
  \zeta(\bm{\hat{p}}(U)) = z_1/z_2.
\end{align}
For a time-optimal solution, $z_1$ and $z_2$ are given by
Eqs.~(\ref{eqn:z1}) and (\ref{eqn:z2}).
We substitute these expressions into
Eq.~(\ref{eqn:stereographic-hopf}) to obtain
$\zeta(t) = f(e^{2i\alpha t})$,
where
\begin{align}
  \label{eqn:mobius}
  f(z) \equiv \frac{e^{-i\phi_0} (z - 1)}{\beta_+ z + \beta_-}.
\end{align}
The function $f(z)$ is a M\"{o}bius transformation.
Since $e^{2i\alpha t}$ describes a circle in the complex plane, and
both stereographic projection and M\"{o}bius transformations preserve
circles, it follows that the time-optimal solutions project to
circular paths on $S^2$.
In Fig.~\ref{projection} we plot example paths for the time-optimal
synthesis of the transformation
$V = e^{i\eta\sigma_z/2}$, which describes a spatial rotation with
axis $\bm{\hat{z}}$ and angle $\eta$.
The paths begin and end at the north pole.
For the paths shown we take $\phi_0 = 0$; alternative paths that also
synthesize $V$ can be obtained by taking different values of $\phi_0$,
and for such paths Fig.~\ref{projection} is rotated about the
$\bm{\hat{z}}$ axis through an angle $\phi_0$.
Under the fiber bundle interpretation, the time-optimal solutions can
be obtained by lifting the circular paths from $S^2$ to $SU(2)$, where
the lifts are performed relative to the connection induced by the
constraint $\bm{L}_z(\bm{r})\cdot\dot{\bm{r}} = 0$.
Another way to visualize the time-optimal solutions is to
stereographically project from the north pole, in which case the
time-optimal solutions map to straight lines on the complex plane.

We have shown that time-optimal solutions project to circular paths on
$S^2$.
We will now show that the length of the path on $S^2$ is equal to
twice the amount of time needed to synthesize the corresponding
transformation.
We first assign coordinates $(\psi, \theta, \phi)$ to an arbitrary
$SU(2)$ transformation $U$ by performing an Euler-angle decomposition:
\begin{align}
  \label{eqn:coordinates-euler}
  U =
  e^{i\psi\sigma_z/2} e^{i\theta\sigma_y/2} e^{i\phi\sigma_z/2}.
\end{align}
We call these coordinates Euler coordinates.
Note that
\begin{eqnarray}
  \bm{\hat{p}}(U) =
  \sin\theta\cos\phi\,\bm{\hat{x}} +
  \sin\theta\sin\phi\,\bm{\hat{y}} +
  \cos\theta\,\bm{\hat{z}},
\end{eqnarray}
so the coordinates $(\theta, \phi)$ are the spherical-polar
coordinates of the point $\bm{\hat{p}}(U)$ on $S^2$.
From Eqs.~(\ref{eqn:coordinates-embedding}),
(\ref{eqn:coordinates-complex}) and (\ref{eqn:coordinates-euler}), it
follows that the Euler coordinates are related to the complex
coordinates $(z_1, z_2)$ and the embedding coordinates
$\bm{r} = (w, x, y ,z)$ by
\begin{align}
  \label{eqn:transformations}
  z_1 &= w + i z = e^{i(\psi+\phi)/2} \cos\theta/2, \\
  z_2 &= y + i x = e^{i(\psi-\phi)/2} \sin\theta/2.
\end{align}
Let us consider a small segment $[t, t+dt]$ of a time-optimal path on
$S^3$.
From Eq.~(\ref{eqn:transformations}), it follows that the arc length
$dt$ of the segment is given by
\begin{align}
  dt &= (d\bm{r}\cdot d\bm{r})^{1/2} \nonumber \\
  &=
  \label{eqn:metric-s3}
 (1/2)(d\theta^2 + d\phi^2 + d\psi^2 +
  2\cos\theta\,d\phi\,d\psi)^{1/2}.
\end{align}
Recall that time-optimal paths satisfy the constraint
$\bm{L}_z(\bm{r})\cdot \dot{\bm{r}} = 0$.
From Eqs.~(\ref{eqn:Lz}) and (\ref{eqn:transformations}), it follows
that in Euler coordinates this constraint takes the form
\begin{align}
  \label{eqn:constraint-euler}
  d\psi + \cos\theta\,d\phi = 0.
\end{align}
We substitute Eq.~(\ref{eqn:constraint-euler}) into
Eq.~(\ref{eqn:metric-s3}) to obtain
\begin{align}
  \label{eqn:t-s}
  dt^2 = (1/4)(d\theta^2 + \sin^2\theta\,d\phi^2) = (1/4)\,ds^2,
\end{align}
where $ds^2$ is the standard metric on $S^2$, which is induced by
the Euclidean metric on $\mathbbm{R}^3$ via the embedding of $S^2$
into $\mathbbm{R}^3$.
From Eq.~(\ref{eqn:t-s}), it follows that the time needed to
synthesize an $SU(2)$ transformation is equal to half the length of
the corresponding path in $S^2$.

\section{Bias field}
\label{sec:bias-field}

Let us now generalize the control problem described in
Sec.~\ref{sec:model-control-problem} by adding a constant bias
magnetic field along the $\bm{\hat{z}}$ axis.
The Hamiltonian for the system is now given by
\begin{align}
  \label{eqn:H-bias}
  H = -\bm{\sigma}\cdot\bm{\hat{n}} + b \sigma_z,
\end{align}
where $b$ characterizes the strength of the bias field.
As before, we assume that $\bm{\hat{n}}$ is constrained to lie in the
$x-y$ plane and thus has the form
$\bm{\hat{n}} = \cos\phi\,\bm{\hat{x}} + \sin\phi\,\bm{\hat{y}}$.
We assume that we are given a target $SU(2)$ transformation $V$ and
bias field value $b$, and we would like to determine the
time dependence of $\phi$ and total evolution time $t$ so as to
synthesize $V$ in a time-optimal fashion.

It is convenient to work in the interaction picture.
We express the Hamiltonian as $H = H_0 + H_i$, where
$H_0 = b \sigma_z$ is the bare Hamiltonian and
$H_i = -\bm{\sigma}\cdot\bm{\hat{n}}$ is the interaction Hamiltonian,
and we define $U_i = e^{iH_0 t} U$ to be the interaction-picture
evolution operator.
The operator $U_i$ satisfies the Schr\"{o}dinger equation
\begin{align}
  i\dot{U}_i = H_I U_i,
\end{align}
where
\begin{align}
  H_I &=
  e^{iH_0 t} H_i e^{-iH_0 t} = -\bm{\sigma}\cdot\bm{\hat{n}}_I, \\
  \bm{\hat{n}}_I &=
  \bm{\hat{x}} \cos \phi_I + \bm{\hat{y}} \sin \phi_I, \\
  \phi_I &= \phi + 2b t.
\end{align}
From the results of Sec.~\ref{sec:homogeneous-problem}, it follows
that the time-optimal solution for $\phi_I$ is given by
$\phi_I = \phi_0 + \omega t$, where $\phi_0$ and $\omega$ are
constants, and the complex coordinates $(z_1(U_i), z_2(U_i))$ of $U_i$
are given by Eqs.~(\ref{eqn:z1}) and (\ref{eqn:z2}).
Since $U = e^{-iH_0 t} U_i$, it follows that the complex coordinates
$(z_1(U), z_2(U))$ of $U$ are given by
\begin{align}
  \label{eqn:z1-bias}
  z_1(U) &=
  (2\alpha)^{-1} e^{-ib t}
  (\beta_+ e^{i\beta_- t} + \beta_- e^{-i\beta_+ t}), \\
  \label{eqn:z2-bias}
  z_2(U) &=
  (2\alpha)^{-1} e^{-i(\phi_0 + b t)} (e^{i\beta_- t} - e^{-i\beta_+ t}),
\end{align}
where $\alpha$ and $\beta_\pm$ are given by
Eq.~(\ref{eqn:alpha-beta}).
Given the complex coordinates of the target transformation $V$, we can
invert Eqs.~(\ref{eqn:z1-bias}) and (\ref{eqn:z2-bias}) to determine
the parameters needed to synthesize $V$ in a time-optimal fashion.

\section{Inhomogeneous control}
\label{sec:inhomogeneous-problem}

We will now generalize the control problem described in
Sec.~\ref{sec:model-control-problem} to the case of inhomogeneous
control.
We consider an ensemble of $N$ spin-$1/2$ particles, where particle
$i$ is in a magnetic field $\bm{B}_i = B_i \bm{\hat{n}}$ with
magnitude $B_i$ and direction $\bm{\hat{n}}$.
The Hamiltonian for particle $i$ is
\begin{align}
  \label{eqn:H-inhomogeneous}
  H_i = -\chi_i\bm{\sigma}\cdot\bm{\hat{n}},
\end{align}
where $\chi_i \equiv \mu B_i$.
As before, we assume that $\bm{\hat{n}}$ is constrained to lie in the
$x-y$ plane and thus has the form
$\bm{\hat{n}} = \cos\phi\,\bm{\hat{x}} + \sin\phi\,\bm{\hat{y}}$.
We note that the single control parameter $\phi$ governs the evolution
of all $N$ particles.
If we evolve the ensemble for a time $t$ while varying the control
parameter $\phi$, we obtain $SU(2)$ evolution operators
$\{U_1(t), \cdots, U_N(t)\}$, where $U_i(t)$ is the evolution operator
for particle $i$.
We assume that we are given a list of target $SU(2)$ transformations
$\{V_1, \cdots, V_N\}$ and a list of field values
$\{\chi_1, \cdots, \chi_N\}$.
We would like to determine the time dependence of $\phi$ and total
evolution time $t$ such that $U_i(t) = V_i$ for $i=1, \cdots, N$,
and $t$ is as small as possible.

We begin by adapting the formalism developed in
Sec.~\ref{sec:homogeneous-problem} to the case of the Hamiltonian
$H_i$ given in Eq.~(\ref{eqn:H-inhomogeneous}).
We denote the embedding coordinates of $U_i$ by $\bm{r}_i$ and the
complex coordinates of $U_i$ by $(z_{1i}, z_{2i})$.
The Schr\"{o}dinger equation in embedding coordinates is
\begin{align}
  \label{eqn:schrodinger-embedding-inhomogeneous}
  \dot{\bm{r}}_i =
  \chi_i(
  n_x \bm{L}_x(\bm{r}_i) +
  n_y \bm{L}_y(\bm{r}_i) +
  n_z \bm{L}_z(\bm{r}_i)),
\end{align}
From Eq.~(\ref{eqn:schrodinger-embedding-inhomogeneous}) and the
orthonormality of the vector fields $\bm{L}_k$, it follows that the
magnitude of the tangent vector $\dot{\bm{r}}_i$ is
$|\dot{\bm{r}}_i| = \chi_i$, so the arc length $s$ of the path traced
out by $\bm{r}_i$ in $S^3$ is related to the time $t$ by
$s = \chi_i t$.
The Schr\"{o}dinger equation in complex coordinates is
\begin{align}
  \label{eqn:schrodinger-complex-inhomogeneous-1}
  \dot{z}_{1i} &= -i\chi_i e^{-i\phi} z_{2i}^*, \\
  \label{eqn:schrodinger-complex-inhomogeneous-2}
  \dot{z}_{2i} &= i\chi_i e^{-i\phi} z_{1i}^*.
\end{align}
From Eqs.~(\ref{eqn:schrodinger-complex-inhomogeneous-2}) and
(\ref{eqn:schrodinger-complex-inhomogeneous-2})
we obtain the decoupled equations of motion
\begin{align}
  \label{eqn:schrodinger-decoupled-inhomogeneous-1}
  \ddot{z}_{1i} + i\dot{\phi}\dot{z}_{1i} + \chi_i^2 z_{1i} &= 0, \\
  \label{eqn:schrodinger-decoupled-inhomogeneous-2}
  \ddot{z}_{2i} + i\dot{\phi}\dot{z}_{2i} + \chi_i^2 z_{2i} &= 0,
\end{align}
and the identities
\begin{align}
  \label{eqn:conservation-laws-inhomogeneous-1}
  \dot{z}_{1i} z_{1i}^* + \dot{z}_{2i} z_{2i}^* &= 0, \\
  \label{eqn:conservation-laws-inhomogeneous-2}
  \dot{z}_{2i} z_{1i} - \dot{z}_{1i} z_{2i} &= i\chi_i e^{-i\phi}.
\end{align}

We can obtain a time-optimal solution to the control problem by
minimizing the action
\begin{align}
  \label{eqn:action-inhomogeneous}
  S = \sum_i A_i + \sum_{i \neq j} (B_{ij} + C_{ij}),
\end{align}
where
\begin{align}
  A_i &=
  \chi_i \int (
  |\bm{r}_i'| +
  \gamma_i(|\bm{r}_i|^2 - 1) +
  \lambda_i \bm{L}_z(\bm{r}_i) \cdot \bm{r}_i')\,du, \\
  B_{ij} &=
  b_{ij} \int
  (\bm{L}_x(\bm{r}_i) \cdot \bm{r}_i' -
  \bm{L}_x(\bm{r}_j)\cdot \bm{r}_j')\,du, \\
  C_{ij} &=
  c_{ij} \int
  (\bm{L}_y(\bm{r}_i) \cdot \bm{r}_i' -
  \bm{L}_y(\bm{r}_j)\cdot \bm{r}_j')\,du,
\end{align}
and $\gamma_i$, $\lambda_i$, $b_{ij}$, and $c_{ij}$ are Lagrange
multipliers.
The terms $A_i$ are straightforward generalizations of the action
(\ref{eqn:action}) for the original control problem; the prefactor
$\chi_i$ accounts for the fact that the arc length $s$ of a path in
$S^3$ is related to the time $t$ by $s = \chi_i t$.
The terms $B_{ij}$ and $C_{ij}$ impose the constraints
$\bm{L}_x(\bm{r}_i)\cdot\dot{\bm{r}}_i =
\bm{L}_x(\bm{r}_j)\cdot\dot{\bm{r}}_j$ and
$\bm{L}_y(\bm{r}_i)\cdot\dot{\bm{r}}_i =
\bm{L}_y(\bm{r}_j)\cdot\dot{\bm{r}}_j$;
from Eqs.~(\ref{eqn:projection-x}) and (\ref{eqn:projection-y}), we
see that these constraints account for the fact that the same control
parameter $\phi$ governs the evolution of all $N$ evolution operators
$\{U_1, \dots, U_N\}$.

We now follow the same procedure described in
Sec.~\ref{sec:homogeneous-problem}: we write down the Euler-Lagrange
equations for $S$, subtract the decoupled Schr\"{o}dinger equation
(\ref{eqn:schrodinger-complex-inhomogeneous-1}) and
(\ref{eqn:schrodinger-complex-inhomogeneous-2}), and use the
identities
(\ref{eqn:conservation-laws-inhomogeneous-1}) and
(\ref{eqn:conservation-laws-inhomogeneous-2}) to obtain equations that
involve only the Lagrange multipliers and the control parameter
$\phi$.
We find that
\begin{align}
  \label{eqn:multipliers-a1}
  \chi_i^2(2\lambda_i - \dot{\phi}) =
  e^{i\phi}\sum_{ij} (\dot{w}_{ij} - \dot{w}_{ji}), \\
  \label{eqn:multipliers-a2}
  i\dot{\lambda}_i - 2\gamma_i - \chi_i^2 =
  -2i e^{i\phi} \sum_{ij} (w_{ij} - w_{ji}),
\end{align}
where $w_{ij} \equiv b_{ij} + i c_{ij}$.

For the case $N=2$ we can solve Eqs.~(\ref{eqn:multipliers-a1}) and
(\ref{eqn:multipliers-a2}) to obtain an equation of motion for $\phi$.
From Eqs.~(\ref{eqn:multipliers-a1}) it follows that
\begin{align}
  \label{eqn:lambdaA}
  \lambda_1 &= (1/2)(\dot{\phi} + \alpha/\chi_1^2), \\
  \label{eqn:lambdaB}
  \lambda_2 &= (1/2)(\dot{\phi} - \alpha/\chi_2^2),
\end{align}
where
\begin{align}
  \label{eqn:alpha}
  \alpha \equiv \dot{w} e^{i\phi}
\end{align}
and $w \equiv w_{12} - w_{21}$.
From Eqs.~(\ref{eqn:multipliers-a2}) it follows that
\begin{align}
  \label{eqn:multipliers-b1}
  &\dot{\lambda}_1 + \dot{\lambda}_2 = 0, \\
  \label{eqn:w}
  &w =
  -(1/4)(\dot{\lambda}_1 - \dot{\lambda}_2 + 2i\beta) e^{-i\phi},
\end{align}
where
$\beta = 2\gamma_1 + \chi_1^2 = -(2\gamma_2 + \chi_2^2)$.
We integrate Eq.~(\ref{eqn:multipliers-b1}) to obtain
\begin{eqnarray}
  \label{eqn:multipliers-c}
  \lambda_1 + \lambda_2 = A,
\end{eqnarray}
where $A$ is an integration constant.
We solve Eqs.~(\ref{eqn:lambdaA}), (\ref{eqn:lambdaB}), and
(\ref{eqn:multipliers-c}) for $\lambda_1$, $\lambda_2$, and $\alpha$
in terms of $\dot{\phi}$ and $A$:
\begin{align}
  \label{eqn:lambdaA-solve}
  \lambda_1 &= (\chi/2)\dot{\phi} - (\chi'/2\chi_1^2)A, \\
  \label{eqn:lambdaB-solve}
  \lambda_2 &= -(\chi/2)\dot{\phi} + (\chi'/2\chi_2^2)A, \\
  \label{eqn:alpha-solve}
  \alpha &= \chi'(\dot{\phi} - A),
\end{align}
where
\begin{align}
  \chi &\equiv
  \frac{\chi_1^2 + \chi_2^2}{\chi_1^2 - \chi_2^2}, &
  \chi' &\equiv
  \frac{2\chi_1^2\chi_2^2}{\chi_1^2 - \chi_2^2}.
\end{align}
We substitute Eqs.~(\ref{eqn:lambdaA-solve}) and
(\ref{eqn:lambdaB-solve}) for $\lambda_1$ and $\lambda_2$ into
Eq.~(\ref{eqn:w}) to obtain
\begin{align}
  \label{eqn:w-solve}
  w = -(1/4)(\chi\ddot{\phi} + 2i\beta)e^{-i\phi}.
\end{align}
We differentiate Eq.~(\ref{eqn:w-solve}) with respect to time and
substitute the resulting expression for $\dot{w}$ into
Eq.~(\ref{eqn:alpha}) to obtain
\begin{align}
  \label{eqn:alpha-solve2}
  \alpha = -(1/4)(
  \chi\dddot{\phi} + 2i\dot{\beta} -
  i\dot{\phi}(\chi\ddot{\phi} + 2i\beta)).
\end{align}
Taking the real and imaginary parts of Eq.~(\ref{eqn:alpha-solve2}),
we find that
\begin{align}
  \label{eqn:multipliers-d1}
  \alpha &= -(1/4)(\chi\dddot{\phi} + 2\beta\dot{\phi}), \\
  \label{eqn:multipliers-d2}
  0 &= -(1/4)(2\dot{\beta} - \chi\ddot{\phi}\dot{\phi}).
\end{align}
We integrate Eq.~(\ref{eqn:multipliers-d2}) to obtain
\begin{align}
  \label{eqn:beta-solve}
  \beta = (\chi/4)\dot{\phi}^2 + B,
\end{align}
where $B$ is an integration constant.
Substituting Eqs.~(\ref{eqn:alpha-solve2}) for $\alpha$ and
(\ref{eqn:beta-solve}) for $\beta$ into Eq.~(\ref{eqn:multipliers-d1}),
we find that
\begin{align}
  \dddot{\phi} + \dot{\phi}^3/2 + (2B/\chi)\dot{\phi} +
  (4\chi'/\chi)(\dot{\phi} - A) = 0.
\end{align}
So the control parameter $\phi$ satisfies the equation of motion
\begin{align}
  \label{eqn:solve-phi-inhomogeneous}
  \dddot{\phi} + \dot{\phi}^3/2 + b\dot{\phi} + a = 0,
\end{align}
where
$a \equiv -(4\chi'/\chi)A$ and
$b \equiv (2/\chi)B - 4\chi'/\chi$.
We note that since the integration constants $A$ and $B$ can take any
values, the parameters $a$ and $b$ can also take any values, and are
thus not constrained by the values of $\chi_1$ and $\chi_2$.

Given initial conditions ($\phi_0, \dot{\phi}_0, \ddot{\phi}_0)$ and
parameters $(a, b)$, we can integrate
Eq.~(\ref{eqn:solve-phi-inhomogeneous}) to obtain a time-optimal
solution for $\phi$.
Given this time-optimal solution, we can integrate the Schr\"{o}dinger
equations
(\ref{eqn:schrodinger-decoupled-inhomogeneous-1}) and
(\ref{eqn:schrodinger-decoupled-inhomogeneous-2})
subject to the initial conditions $(z_{1i}, z_{2i}) = (1,0)$ to obtain
the complex coordinates of a pair of evolution operators
$\{U_1, U_2\}$.
It is useful to view the parameters
$(\phi_0, \dot{\phi}_0, \ddot{\phi}_0, a, b, t)$ as a generalization
of the time-optimal coordinates described in
Sec.~\ref{sec:homogeneous-problem}.
The two integrations then define a coordinate transformation from the
time-optimal coordinates to the complex coordinates of the pair of
evolution operators $\{U_1, U_2\}$.
Given target $SU(2)$ transformations $\{V_1, V_2\}$ and field values
$\{\chi_1, \chi_2\}$, we can write down the complex coordinates of
$\{V_1, V_2\}$ and then invert this coordinate transformation to
determine the time dependence of the control parameter $\phi$ and the
total evolution time $t$ needed to synthesize $V_1$ and $V_2$ in a
time-optimal fashion.
We have thus formally solved the inhomogeneous control problem for the
case $N=2$.

We note that the parameters
$(\phi_0, \dot{\phi}_0, \ddot{\phi}_0, a, b)$ determine a time-optimal
evolution for the control parameter $\phi$, and this evolution,
together with the parameters $(t, \chi_1, \chi_2)$, determine a pair
of evolution operators $\{U_1, U_2\}$.
It is interesting that the time-optimality of $\phi$ does not depend
on the field values $\chi_1$ and $\chi_2$.
That is, if we hold the time dependence of $\phi$ and the total
evolution time $t$ fixed, and vary $\chi_1$ and $\chi_2$, we will
synthesize different evolution operators $U_1$ and $U_2$, but it will
always be the case that the synthesis of these operators is
time-optimal.

\begin{figure}[t]
  \centering
  \includegraphics{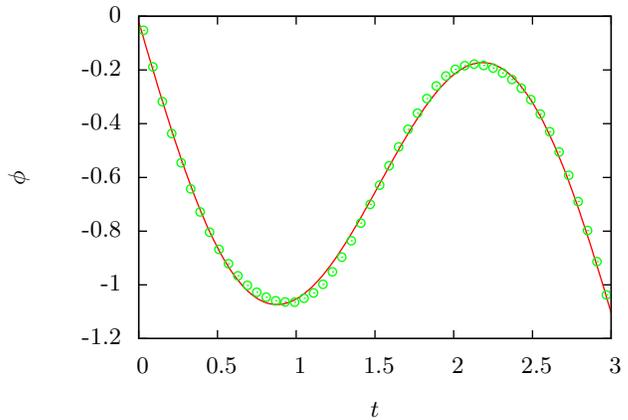}
  \caption{
    \label{fig:phi}
    (Color online)
    Control parameter $\phi$ versus time $t$.
    The solid curve is obtained by numerically integrating
    Eq.~(\ref{eqn:solve-phi-inhomogeneous}); the points are obtained
    from a numerical gradient-ascent search with $t=3$ and $R=50$.
  }
\end{figure}

Let us now consider a specific example.
We will take the field values to be $\chi_1 = 1/2$ and
$\chi_2 = 3/2$, and consider the pair of transformations
$\{V_1, V_2\}$
whose time optimal coordinates are
$\phi_0 = 0$, $\dot{\phi}_0=-2$, $\ddot{\phi}_0=0$, $a=2$, $b=3$,
$t=3$.
We numerically integrate the equation of motion
(\ref{eqn:solve-phi-inhomogeneous}) to determine the time evolution of
the control parameter $\phi$ that synthesizes $V_1 = U_1(t)$ and
$V_2 = U_2(t)$ in a time-optimal fashion, and we numerically
integrate the Schr\"{o}dinger equations
(\ref{eqn:schrodinger-complex-inhomogeneous-1}) and
(\ref{eqn:schrodinger-complex-inhomogeneous-2}) to determine the
complex coordinates of the pair $\{V_1, V_2\}$.
In Fig.~\ref{fig:phi} we plot the resulting time-optimal evolution of
$\phi$.

\begin{figure}[t]
  \centering
  \includegraphics{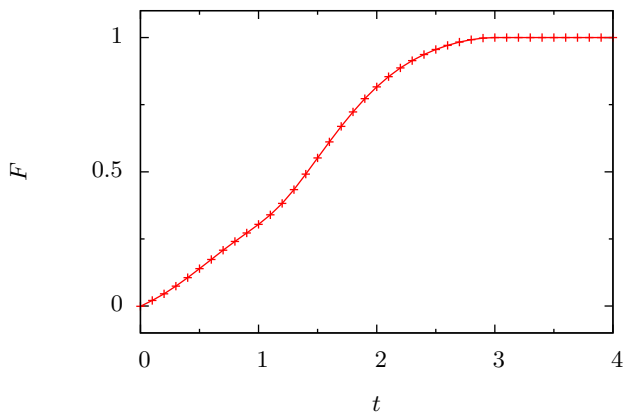}
  \caption{
    \label{fig:F}
    (Color online)
    Maximum fidelity $F_{max}$ versus time $t$, as determined by a
    gradient-ascent search with $R=50$.
  }
\end{figure}

We verify that the synthesis of $V_1$ and $V_2$ is time optimal as
follows.
Given arbitrary $SU(2)$ transformations $A_1$ and $A_2$, we define the
fidelity with which $A_1$ and $A_2$ approximate $V_1$ and $V_2$ to be
\begin{align}
  \label{eqn:F}
  F =
  (1/4)(\textup{Tr}[V_1^\dagger A_1] + \textup{Tr}[V_2^\dagger A_2]).
\end{align}
The fidelity ranges from $-1$ to $1$, where $F=1$ if $A_1 = V_1$ and
$A_2 = V_2$, and $F$ decreases as the deviation of $A_1$ and $A_2$
from $V_1$ and $V_2$ increases.
We fix the total evolution time $t$, and we discretize the
time evolution of the control parameter by dividing $t$ into $R$
timesteps of duration $\delta t = t/R$.
We define $\phi_r = \phi(r \delta t)$ to be the value of the control
field at timestep $r$.
We then take $A_1 = U_1(t)$ and $A_2 = U_2(t)$, and perform a
numerical gradient-ascent search to maximize $F$ with respect to the
discretized control parameter values
$\{\phi_0, \cdots, \phi_{R-1}\}$.
In Fig.~\ref{fig:F} we plot the numerically-determined maximum
fidelity $F_{max}$ as a function of $t$ for $R=50$.
Since $F_{max}$ first reaches $1$ at $t=3$, we see that the evolution
described above is indeed time-optimal.
In Fig.~\ref{fig:phi}, we plot the time-optimal evolution of $\phi$
for $t=3$, as determined by the gradient-ascent search.
We find good agreement with the time-optimal evolution of $\phi$
obtained by integrating the equation of motion
(\ref{eqn:solve-phi-inhomogeneous}).

\section{Summary}

We have considered a quantum control problem involving a spin-$1/2$
particle in a magnetic field.
We have analytically solved for the time dependence of the control
parameter needed to synthesize an arbitrary $SU(2)$ transformation in
a time-optimal fashion, and we have generalized our solution to the
case of an inhomogeneous control problem involving an ensemble of
spin-$1/2$ systems.

\section{Acknowledgements}

The author would like to thank Ivan Deutsch for valuable discussions
and suggestions.
This research was supported by NSF Grant No. PHY-0903953.

\end{document}